\documentclass{IEEEtran4PSCC}
\ifCLASSINFOpdf
   \usepackage[pdftex]{graphicx}
\else
   \usepackage[dvips]{graphicx}
\fi
\usepackage[cmex10]{amsmath}
\makeatletter
\let\old@ps@headings\ps@headings
\let\old@ps@IEEEtitlepagestyle\ps@IEEEtitlepagestyle
\def\psccfooter#1{%
    \def\ps@headings{%
        \old@ps@headings%
        \def\@oddfoot{\strut\hfill#1\hfill\strut}%
        \def\@evenfoot{\strut\hfill#1\hfill\strut}%
    }%
    \def\ps@IEEEtitlepagestyle{%
        \old@ps@IEEEtitlepagestyle%
        \def\@oddfoot{\strut\hfill#1\hfill\strut}%
        \def\@evenfoot{\strut\hfill#1\hfill\strut}%
    }%
    \ps@headings%
}
\makeatother

\psccfooter{%
        \parbox{\textwidth}{\hrulefill \\ \small{22nd Power Systems Computation Conference} \hfill \begin{minipage}{0.2\textwidth}\centering \vspace*{4pt} \includegraphics[scale=0.06]{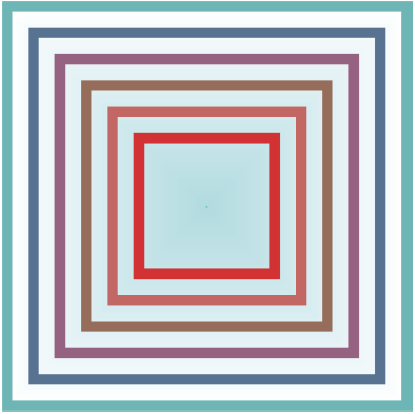}\\\small{PSCC 2022} \end{minipage} \hfill \small{Porto, Portugal --- June 27 -- July 1, 2022}}%
}

\usepackage{amsfonts}
\usepackage{amssymb}
\usepackage{MnSymbol}%

\usepackage{xcolor}
\usepackage{bm}
\usepackage{algorithm}
\usepackage{algpseudocode}

\makeatletter
\newenvironment{protocol}[1][htb]{%
    \renewcommand{\ALG@name}{Protocol}
   \begin{algorithm}[#1]%
  }{\end{algorithm}}
\makeatother

\begin{document}

\title{Distributed Privacy-Preserving Electric Vehicle Charging  Control Based on Secret Sharing}

\author{
\IEEEauthorblockN{Xiang Huo and Mingxi Liu}
\IEEEauthorblockA{Department of Electrical and Computer Engineering \\
University of Utah\\
Salt Lake City, UT, USA\\
\{xiang.huo, mingxi.liu\}@utah.edu}
}

\maketitle

\begin{abstract}
 
Cooperative electric vehicle (EV) charging control has emerged as a key component in grid-edge resource (GER) management. However, customers' privacy remains a major barrier to large-scale implementation of EV charging control. In this paper, we develop a distributed privacy-preserving EV charging control protocol based on secret sharing (SS) that 1) achieves scalability over EV population size; 2) enjoys higher computation efficiency compared to homomorphic encryption (HE) based methods; and 3) secures the privacy of the participating EVs against honest-but-curious adversaries and external eavesdroppers. The cooperative EV charging control is formulated to achieve overnight valley-filling and framed into the projected gradient algorithm (PGA) structure as a distributed optimization problem. SS is integrated into PGA to achieve secure updates of both primal and dual variables. Theoretical security analyses and simulations in a residential area are conducted to prove the privacy preservation guarantee as well as the efficacy and efficiency of the proposed privacy preservation method. Broadly, the proposed method can be readily extended to various GER control applications.

\end{abstract}



\section{Introduction}

The increasing penetration of electric vehicles (EV) has brought unprecedented challenges and opportunities to the power grid. EVs are more economic and environmental friendly owing to their petroleum independence and reduced greenhouse gas emissions \cite{nanaki2013comparative}. Nevertheless, without proper coordination and control, the injection of enormous EV charging loads can easily jeopardize the stability of the distribution grid. On the other side, as already proven \cite{ovalle2015optimal,wang2013grid}, EV charging load coordination and regulation can provide a range of grid services, e.g., peak shaving and valley-filling. Hence, a commensurate EV charging control framework is demanded to attenuate the impacts and promote the benefits.

To improve the scalability and resilience of the distribution grid, distributed controller and optimization strategies for large-scale EV charging problems are drawing more attentions \cite{nimalsiri2019survey}. Faddel \emph{et al.} in \cite{faddel2018automated} focused on demand side management programs and designed a distributed controller to  coordinate the charging behaviors of EVs. In \cite{RiveraDistributed}, an optimization framework based on the alternating directions method of multipliers (ADMM) was proposed to achieve computational scalability of large-scale EV charging control. The proposed method can concurrently solve the   valley-filling and cost minimization optimization problems. Similarly, Shao \emph{et al.} in \cite{shao7524029} proposed a partial decomposition method based on the Lagrangian relaxation for EV charging control, aiming at  reducing the total generation cost and alleviating the grid congestion. To deal with the high penetration of EVs, Ma \emph{et al.} in \cite{ma2014distributed} designed three distributed EV charging control algorithms and compared the trade-off between the feasibility and the optimality under power network capacity constraints. 

However, the abovementioned  distributed EV charging control algorithms as well as general distributed multi-agent optimization techniques require frequent communications between participants, which can lead to potential privacy breaches, e.g., the charging behaviour of an EV can reveal the owner's 
driving patterns \cite{yucel2019efficient}. Therefore, privacy has emerged as a major blocking barrier to the implementation of distributed control.
One popular tool that can be integrated into distributed control to achieve privacy preservation is homomorphic encryption (HE). HE enables secure arithmetic operations over encrypted data, and when decrypted, the result matches operations performed on the original plaintext. Li \emph{et al.} in \cite{li2013eppdr} designed a privacy-preserving demand response scheme based on HE, and an adaptive key evolution was proposed to balance the trade-off between the communication efficiency and security level. In \cite{ruan2019secure}, a privacy-preserving approach based on partial HE was proposed to solve consensus problems securely. The proposed approach leverages cryptography to protect each agent from disclosing its true states to the neighbors and guarantee the  convergence to the consensus value. Brettschneider \emph{et al.} in \cite{brettschneider2016homomorphic} proposed an HE scheme for distributed load adaption, where the complexity of the communication and computational overhead was analyzed. 

Though HE-based methods have guaranteed security and high accuracy, their industrial applicability is severely limited by the high computational complexity induced by encryption and decryption. In contrast to cryptographic strategies,  another non-cryptographic strategy to enabling privacy preservation is secret sharing (SS) which was firstly introduced by Shamir in \cite{shamir1979share}. SS is polynomial based and it deploys a manager to divide a secret message to multiple shares and distribute each share to an agent. In this manner, individuals learn nothing about the secret, but any set of agents that has a cardinality  more than a threshold can reconstruct the secret. Compared to HE-based methods, SS-based strategies remove the computing burden caused by frequent encryption and decryption, and achieve improved efficiency by adopting share computation and secret reconstruction. In \cite{daru2019encrypted}, a distributed cloud-based controller using SS was proposed to achieve encrypted linear state feedback, where the encryption scheme was modified by one-time pad to reduce the computational overhead. Li \emph{et al.} in \cite{li2019privacy}
developed a distributed privacy-preserving  
algorithm based on SS specifically for averaging consensus problems. In \cite{wagh2020distributed}, a distributed privacy-preserving framework based on SS was proposed to protect the privacy of the customers. The designed framework can protect the readings of a smart meter from dedicated aggregators and the electrical utility. 

However, none of the existing SS-based methods are applicable to EV charging control, as they either rely on the coordination of a third party or are designed for special problem formulations that are not suitable for EV charging control. This paper fills this gap by developing a novel SS-based distributed privacy-preserving EV charging control scheme that is third party free and concurrently achieves privacy preservation, scalablity, high computing efficiency, and high accuracy.


The contribution of this paper is three-fold: 1) We, for the first time, integrate SS into distributed optimization algorithms to achieve scalable EV charging control for valley filling. 2) The proposed protocol enjoys high computing efficiency compared to HE-based methods. 3) We investigate the attacks from both internal and external adversaries, and provide theoretically sound proof of security. The proposed method can be extended to other grid-edge resource (GER) controls.

\section{Main Results \& Methodologies}

\subsection{Problem Formulation}
In a distribution feeder, the high electricity demand during daytime and the low  electricity demand at night create a valley in the demand profile. Such valley can increase the grid operation cost like shutting down or restarting large power plants \cite{zhang2014coordinating}. In this section, we formulate an EV charging control optimization problem to fill the overnight demand valley. Suppose $n$ EVs need to be charged during the valley filling period $\tilde{T} = [1:T]$ of length $T\Delta T$ where $\Delta T$ is the sampling period. We denote the charging profile of the $i$th EV by $\bm{x}_i = [x_i(1),\ldots,x_i(T)]^\mathsf{T}\in \mathbb{R}^{T}$ where $x_i(t)$ is a scalar representing its charging rate at time $t$. For the $i$th EV, its charging rate is constrained by 
\begin{equation}
    \bm{0} \leq \bm{x}_{i} \leq \bm{r}_i^u, \label{1s}
\end{equation}
where $\bm{r}_i^u$ denotes the maximum charging rate of the $i$th EV. 

To guarantee all EVs can be charged to the desired energy level by the end of valley-filling period, the summation of the charging loads of the $i$th EV during period $\tilde{T}$ should satisfy 
\begin{equation}
    \bm{G}\bm{x}_i = d_i,\label{2s}
\end{equation}
where $\bm{G} = [\Delta T,\ldots,\Delta T] \in \mathbb{R}^{1\times T}$ denotes the aggregation vector and $d_i$ denotes the charging demand of the $i$th EV. 

Filling the overnight demand valley is equivalent to flattening the total demand curve by exploiting the EV charging loads \cite{gan2012optimal}. Therefore, the valley-filling optimization problem can be formulated into a quadratic programming problem with the abovementioned constraints as 
\begin{equation} \label{s1}
\begin{aligned}
    &\underset{\bm{x}}{\text{min}} \quad {\frac{1}{2}\left\|\bm{P}_{b}+\sum_{i=1}^{n} \bm{x}_i\right\|_{2}^{2}} \\
& \,\, \text{s.t.}  \, \quad \bm{0} \leq \bm{x}_{i} \leq \bm{r}_i^u, \quad \forall i=1,2, \ldots, n, \\
& \,\,\,\, \qquad 
\bm{G}\bm{x}_i = d_i, \quad \forall i=1,2, \ldots, n, 
\end{aligned}
\end{equation}
where $\bm{P}_{b}$ denotes the baseline load. Other network constraints, e.g., 
the voltage limits of the distribution network, can also be included in problem \eqref{s1} and have no impact on privacy preservation. In this paper, to better illustrate the design of the privacy preservation method, we avoid over-complicating the optimization problem by omitting those network constraints. 


To solve the valley-filling problem in \eqref{s1}, $n$   agents (EVs) can work corporately using projected gradient algorithms (PGA) \cite{bertsekas2015parallel}. The PGA based method is widely adopted in distributed and decentralized optimization, e.g., multiuser optimization problem
\cite{koshal2011multiuser}, decentralized EV charging control  \cite{liu2017decentralized}, and convex games \cite{zhu2016distributed}. Using PGA, agent $i$  updates its decision variable by following
\begin{equation} \label{eq1}
    \bm{x}_i^{\ell+1} = \mathbb{P}_{X_i}[\bm{x}_i^{\ell} - \gamma^{\ell}_i \Phi_i (\bm{x}^{\ell})], 
\end{equation}
where $\ell$ denotes the iteration index, $\bm{x} \triangleq [\bm{x}_1^\mathsf{T},\ldots,\bm{x}_n^\mathsf{T}]^\mathsf{T}$, $X_i$ denotes the feasible set of $\bm{x}_i$, $\gamma^{\ell}_i$ is the step size length of agent $i$, $\Phi_i(\cdot)$ is the first-order gradient of the Lagrangian function \emph{w.r.t.} $\bm{x}_i$, and $\mathbb{P}_{X_i}[\cdot]$ denotes the Euclidean projection operation onto set $X_i$. 


To solve \eqref{s1} with any projection-based method, we firstly calculate the relaxed Lagrangian of  \eqref{s1} as
\begin{equation} \label{5}
    \mathcal{L}(\bm{x},\bm{\lambda})= \frac{1}{2}\left\|\bm{P}_{b}+\sum_{i=1}^{n} \bm{x}_i\right\|_{2}^{2}+\sum_{i=1}^{n}\lambda_i(\bm{G}\bm{x}_i - d_i), 
\end{equation}
where  $\bm{\lambda} \triangleq [\lambda_1,\ldots,\lambda_n]^\mathsf{T}$ and $\lambda_i$ is the dual variable associated with the $i$th equality constraint. Therefore, the subgradients of $\mathcal{L}(\bm{x},\bm{\lambda})$ \emph{w.r.t.} $\bm{x}_i$ and $\lambda_i$ can be readily obtained as
\begin{subequations}\label{6s}
\begin{align}
    \nabla_{x_i} \mathcal{L}(\bm{x},\bm{\lambda}) & =  \bm{P}_{b}+\sum_{i=1}^{n} \bm{x}_i  + \lambda_i{\bm{G}}^\mathsf{T}, \label{4as}\\
     \nabla_{\lambda_i} \mathcal{L}(\bm{x},\bm{\lambda}) & = \bm{G}\bm{x}_i - d_i.    \label{4bs}
\end{align}
\end{subequations}

Then, the decision variable (primal variable) and the dual variable can be updated  using the PGA. Herein, we derive the update rule \eqref{11a} for the primal variable using the subgradient in \eqref{4as} and PGA equation in \eqref{eq1}. The dual variable can be updated using subgradient ascent with the subgradient in \eqref{4bs}. Consequently, we have
\begin{subequations} \label{update}
\begin{align}
\bm{x}_{i}^{\ell+1} &=
\Pi_{\mathbb{X}_{i}}\left( \bm{x}_{i}^{\ell}-\gamma_{i} \nabla_{x_{i}} \mathcal{L}\left(\bm{x}^{\ell}, \bm{\lambda}^{\ell}\right)\right), \label{11a}\\
\lambda_i^{\ell+1} &=  
\lambda_i^{\ell}+\beta \nabla_{\lambda_i} \mathcal{L}\left(\bm{x}^{\ell}, \bm{\lambda}^{\ell}\right), \label{11b}
\end{align}
\end{subequations}
where $\mathbb{X}_{i}\triangleq \{\bm{x}_{i} ~|~ \boldsymbol{0} \leq \bm{x}_{i} \leq \bm{r}_i^u \}$ denotes the feasible set of the charging rate limits, and $\gamma_i$ and $\beta$ denote the primal and dual step sizes, respectively.

\noindent \textbf{Remark 1:} $\mathbb{X}_i$ is chosen to only include \eqref{1s} so that \eqref{2s} can mimic the global network constraints in $\mathcal{L}(\bm{x},\bm{\lambda})$. \hfill $\square$


Note that in \eqref{4as}, the calculation of $\nabla_{x_i} \mathcal{L}(\bm{x},\bm{\lambda})$ depends on $\bm{x}_i$'s from all the agents. Therefore, the update in  \eqref{11a} requires decision variable exchanges between agents, which can lead to potential privacy breaches \cite{zhang2018enabling}. To overcome this privacy challenge, in this paper, we aim at designing a novel protocol based on SS that can protect agents' privacy. Specifically, we will design a secure information exchange protocol in which the agents aggregate and distribute information privately without exposing privacy to others. In the following, we first illustrate the standard steps of SS, then develop the SS-based privacy-preserving protocol.

\subsection{Shamir's Secret Sharing Scheme}
\label{chapter_SS}

The essential idea of SS is Lagrange polynomial interpolation, i.e., given $k$ points $(a_1,b_1),\ldots,(a_k,b_k)$ with distinct $a_i$'s on the 2-D plane, there exists a unique polynomial $q(z)$ of degree $k-1$ such that $q(a_i) = b_i, \forall i = 1,\ldots,k$. Any number of points that are strictly smaller than $k$ will not reveal any information about the polynomial, and at least $k$ points are required to reconstruct the polynomial.

In the following, we present a case where a manager wants to distribute an integer secret $s$ to $n$ agents, and at least $k$ agents are required to cooperatively retrieve the secret. The secret is randomized via a polynomial and sent to the agents in the form of shares. Any $\hat{k}$ agents, where $k \leq \hat{k} \leq n$, can reconstruct the secret, otherwise no information about the secret can be revealed. This is called a $(k,n)$-threshold SS which comprises of three steps as 

\begin{enumerate}
    \item \textit{Polynomial Generation:} The manager constructs a random polynomial 
    \begin{equation}
        f(z) = s + c_1z + \cdots + c_{k-1}z^{k-1} \label{f(z)},
    \end{equation}
    where $s$ denotes the secret, the coefficients $c_1,\ldots,c_{k-1}$
    are randomly chosen from a uniform distribution in the integer field $\mathbb{E} \triangleq [0,e)$, where $e$ denotes a prime number that is larger than $s$.

    \item \textit{Share Distribution:} 
    The manager computes the shares with a non-zero integer input and the corresponding output, e.g., set $i=1,\ldots,n$ to retrieve $(i,f(i))$. Then, it distributes the share $f_i$ to the $i$th agent, where
   \begin{equation}
        f_i = f(i) \bmod e \label{f_i}.
    \end{equation}

    \item \textit{Secret Reconstruction:} Since each agent is given a point and $k$ points are sufficient to reconstruct the polynomial based on Lagrange interpolation, any $\hat{k}$ agents, where $k \leq \hat{k} \leq n$, can calculate the secret $s$ using interpolation as
    \begin{equation}
       s =  \sum_{i=1}^{k}f_i\prod_{j=0  \atop j \neq i}^{k} \frac{j}{j-i}.
       \label{recosntruction}
    \end{equation}

\end{enumerate}

Note that the constant term $s$ in \eqref{f(z)}
is exactly the secret that can be calculated by $f(0)=s$. The Shamir's SS scheme requires integers and finite field arithmetic
to guarantee the perfect security, i.e., the coefficients $c_1,\ldots,c_{k-1} $ and the secret $s$ have to be within the field $\mathbb{E}$. Besides, the modular operation in \eqref{f_i} also guarantees that the outputs of the polynomials are within the field $\mathbb{E}$. In \eqref{recosntruction}, though we only calculate the secret $s$, it is also possible to reconstruct  \eqref{f(z)}. But since we are only interested in retrieving the secret, \eqref{recosntruction} is adopted to reduce the unnecessary computing cost.

\subsection{Real Number and Integer Transformation}

 Note that the SS scheme requires modular arithmetic within the field $\mathbb{E}$. However, distributed optimization genetically requires real number calculations, i.e., real decision variables and parameters. Therefore, to integrate SS into distributed optimization, a real number needs to be transformed into an integer. Herein, we formulate the real number to integer transformation rule as 
 \begin{equation}
     z_e = \lfloor10^{\delta} \theta \rfloor,
     \label{11ss}
 \end{equation}
where $\theta$ denotes any real number, $\delta$ denotes the preserved decimal fraction digits, $\lfloor \cdot \rfloor$ denotes the floor operation, and $z_e$ is the transformed integer.
To proceed with the SS, we further need to map $z_e$ to the field $\mathbb{E}$, which can be achieved by 
 \begin{equation}
     z_e^{+} = z_e \bmod e,\label{12ss}
 \end{equation}
where $z_e^{+} \in \mathbb{E}$ denotes the integer that is mapped to the field $\mathbb{E}$. To transform an integer back to the real number, we employ the inverse function $\phi(\cdot)$ that is  defined as 
\begin{align}
\phi(z) = \begin{cases}
     z-e, &\text { if } z \geq  \frac{e}{2}, \\ 
    z, &\text { otherwise. }
\end{cases}
    \label{12}
\end{align}
Therefore, a number in the field $\mathbb{E}$ can be transformed to the original signed integer. Finally, we can simply divide the signed integer by $10^{\delta}$ to convert it to the original real number.

\subsection{Proposed Privacy-Preserving Algorithm}

In this section, we develop a novel privacy-preserving protocol based on SS. During the iterative computations, agents update their primal and dual variables by following  \eqref{update}. The agents can perform the dual update in \eqref{11b} independently without having to communicate with each other. Therefore, the dual variables $\lambda_i$'s can be kept private to the $i$th agents directly. However, the primal update in \eqref{11a} requires the aggregation $\sum_{i=1}^{n}\bm{x}_i$ in \eqref{4as}, leading to mandatory exchange of decision variables between all the agents. To eliminate the potential decision variable leakage caused by the information exchange, we integrate SS into \eqref{update} to achieve the secure message aggregation. 

To clearly state the privacy issues concerned in this paper, we herein define the privacy as follows.

\noindent \textbf{Definition 1:} (\textit{Privacy}) For the $i$th EV, the private information includes its charging profile $\bm{x}_i$, dual variable $\lambda_i$, the maximum charging rates $\bm{r}_i^u$, demand $d_i$, and the primal and dual step sizes $\gamma_i$ and $\beta$. \hfill $\blacksquare$

The privacy in Definition 1 can be categorized into three aspects: 1) private parameters of the EVs (agents), i.e., $\bm{r}_i^u$  and $d_i$; 2) intermediate iteration variables of the agents, i.e., $\bm{x}_i^{\ell}$ and $\lambda_i^{\ell}$; and 3) the parameters used by the PGA, i.e., $\gamma_i$ and $\beta$. Though Definition 1 specifically targets at EV charging control problem, it can also be extended to various
optimization problems in other fields (as illustrated in Remark 3). Besides, the private parameters of the agents can be extended as well, e.g., constraints on charging ramp rate.  

To proceed with the protocol design,  the following assumption is required for the network communication.

\noindent \textbf{Assumption 1:} The network is interconnected through communication channels. \hfill $\blacksquare$

Under Assumption 1, each agent is able to communicate with any other agents. This interconnected communication network can be easily set up, e.g., EVs can communicate in the cloud or on a specific server with internet access \cite{chynoweth2014smart}. In the following proposed framework, we admit that Assumption 1 may introduce an overall high communication load between agents, so our future research direction will target at lifting Assumption 1 to reduce the communication cost, e.g., using a spanning-tree communication network. 

Recall that in the SS scheme, a manager is required to generate the polynomial and distribute the shares. To eliminate the need of a manager, we empower each agent to act as both manager and agent. Each agent can construct polynomials, distribute shares, and receive messages from others.  The $i$th agent firstly generates a set of random parameters $c_{i1}^{\ell},\ldots,c_{ik}^{\ell} \in \mathbb{E}$, then constructs a polynomial as 
\begin{equation}
    p_i^{\ell}(z) = s_i^{\ell} + c_{i1}^{\ell}z + \cdots + c_{ik}^{\ell}z^{k},
    \label{11s}
\end{equation}
where $s_i^{\ell}$ is the secret held by the $i$th agent, i.e., the charging profile $\bm{x}_i^{\ell}$. Note that the charging profiles $\bm{x}_i^{\ell}$'s are real number vectors, while $s_i^{\ell}$'s need to be integers. To resolve this issue, agents should firstly transform secret $\bm{x}_i^{\ell}$ into integer vectors using \eqref{11ss}, then deal with the integer vectors elementwisely.

Having the polynomial constructed, the $i$th agent then evaluates $p_i^{\ell}(\alpha_j), \forall j=1,\ldots,n$ where $\alpha_j$'s are the shared input knowledge among all the agents. Following that, the $i$th agent sends $p_i^{\ell}(\alpha_j), j \neq i$, to the $j$th agent and keeps $p_i^{\ell}(\alpha_i)$ to itself.  After receiving all the messages, agent $i$ is ready to compute the summation of the received messages as 
\begin{equation}
   v_{i}^{\ell} = \sum_{l=1}^{n}p_i^{\ell}(\alpha_l). 
   \label{12s}
\end{equation}

Following the summation, agent $i$ broadcasts $v_{i}^{\ell}$ to other agents without revealing any true self-relevant information. Note that the summation $v_{i}^{\ell}$'s calculated by the agents in \eqref{12s} correspond to the outputs of the polynomial 
\begin{equation}
 \Tilde{p}^{\ell}_i(z) =  \sum_{l=1}^{n} s_l^{\ell} + \sum_{j=1}^{k} (\sum_{l=1}^{n}c_{lj}^{\ell})z^j,
 \label{13s}
\end{equation}
at $z=\alpha_l, \forall l=1,\ldots,n$. The polynomial $ \Tilde{p}_i^{\ell}(z)$ has a degree of $k$, affiliated with the constant term $\sum_{l=1}^{n} s_l^{\ell}$ and the coefficients  $\sum_{l=1}^{n}c_{lj}^{\ell}, j=1,\ldots,k$.
 
Therefore, the $i$th agent can collect in total $n$ points, i.e., $(\alpha_i,v_j^{\ell}),\forall j=1,\ldots,n$, that correspond to the input and output pairs of the polynomial in \eqref{13s}. In the following, each agent is ready to reconstruct the constant term $\sum_{l=1}^{n} s_l^{\ell}$ according to the secret reconstruction step in \eqref{recosntruction}. After obtaining the summation of the secrets $\sum_{l=1}^{n} s_l^{\ell}$, one can apply the inverse function $\phi(\cdot)$ defined in \eqref{12} to acquire the signed integer summation $10^{\delta}\sum_{l=1}^{n} \hat{\bm{x}}_l^{\ell}$ elementwisely. Note that $\hat{\bm{x}}_l^{\ell}$ is a real vector that is approximated by keeping $\delta$ digits of $\bm{x}_l^{\ell}$. Then, the signed integer summation $\sum_{l=1}^{n} \hat{\bm{x}}_l^{\ell} \cdot 10^{\delta}$ can be transformed back to the real vectors by the division of $10^{\delta}$. 

Finally, the agents can update the primal and dual variables according to \eqref{update}. The agents firstly calculate the subgradients defined in \eqref{6s} using the summation $\sum_{l=1}^{n} \hat{\bm{x}}_l^{\ell}$. The $i$th agent then updates its primal variable $\bm{x}_i^{\ell} \rightarrow \bm{x}_i^{\ell+1}$ and  dual variable $\lambda_i^{\ell} \rightarrow \lambda_i^{\ell+1}$.
The convergence error $\epsilon_i^{\ell}$ can be calculated as the summation of the Euclidean distances of the primal and dual variables in two consecutive iterations as
\begin{equation}
    \epsilon_i^{\ell} = 
    \| \bm{x}_i^{\ell+1}-\bm{x}_i^{\ell}\|_2^2 +
    \| \lambda_i^{\ell+1}-\lambda_i^{\ell}\|_2^2.
\end{equation}
EV $i$ can stop the iterations and keep the previous values if $\epsilon_i^{\ell} \leq \epsilon_0$ where $\epsilon_0$ denotes the error tolerance. 

\begin{figure}[!htbp]
    \centering
 \includegraphics[width=0.45\textwidth,trim = 0mm 0mm 0mm 0mm, clip]{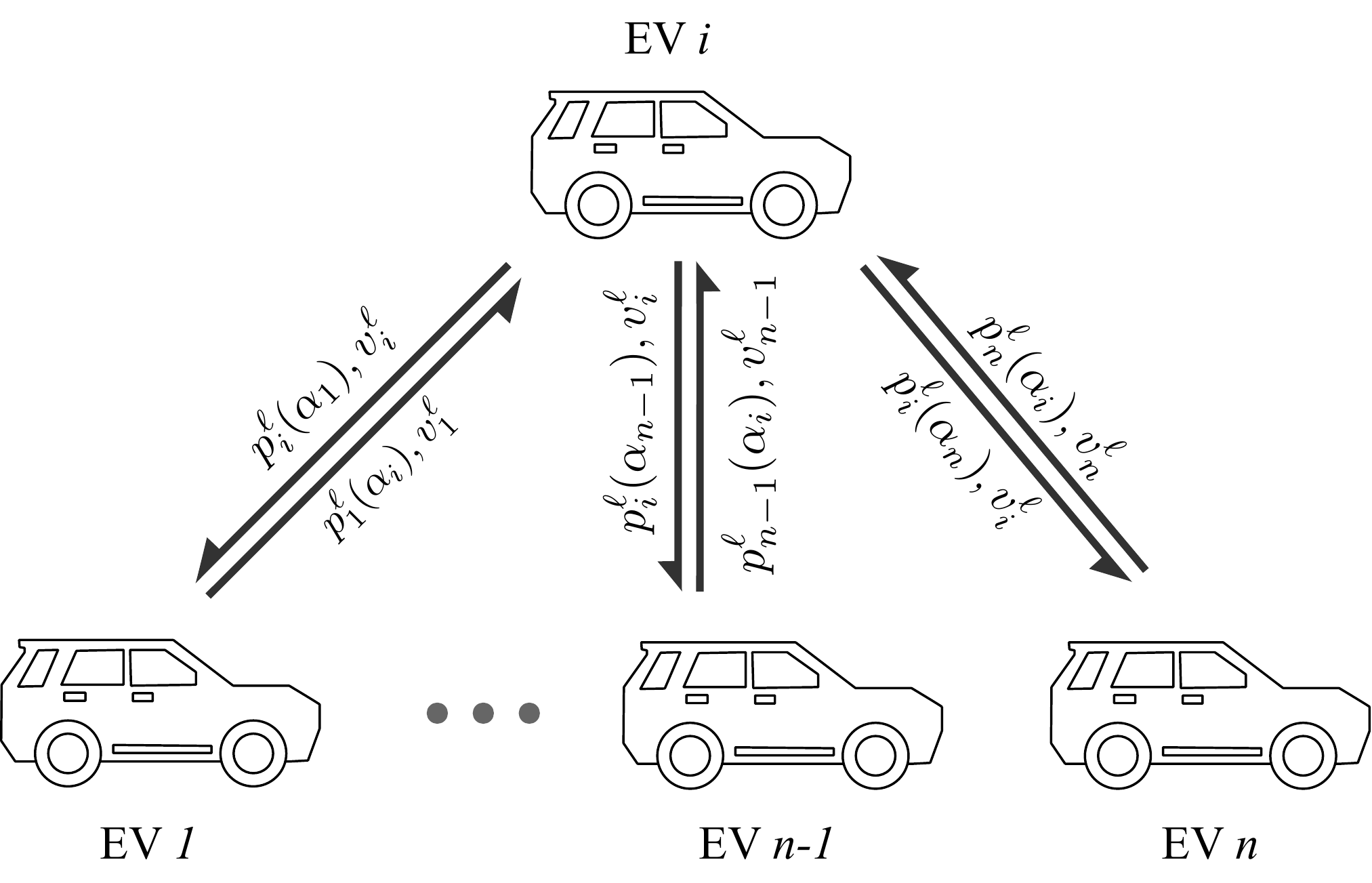}
    \caption{The information exchange between the $i$th EV and other EVs.}
    \label{information}
\end{figure}

The detailed procedure of the proposed method is presented via Protocol \ref{protocol_1}. The information exchange between the $i$th EV and other EVs is shown in Fig. \ref{information}.
\begin{protocol} 
\caption{Distributed SS-based  privacy-preserving EV charging control protocol}
\begin{algorithmic}[1]

\State The $i$th EV is associated with an integer $\alpha_i$, and $\alpha_j,\forall j=1,\ldots,n$, are common knowledge among all EVs.

\State All EVs initialize primal and dual variables, tolerance $\epsilon_0$, iteration counter $\ell=0$, and maximum iteration $\ell_{max}$.

\While{ $\epsilon_i^{\ell} > \epsilon_0$ and $\ell < \ell_{max}$}

\State EV $i$ generates a set of random parameters satisfying  $c_{i1}^{\ell},\ldots,c_{ik}^{\ell} \in \mathbb{E}$, then constructs the polynomial in \eqref{11s}. 

\State The $i$th EV calculates $p_i^{\ell}(\alpha_j), \forall j=1,\ldots,n$, and sends $p_i^{\ell}(\alpha_j), j \neq i$ to the $j$th EV and keeps $p_i^{\ell}(\alpha_i)$ to itself.

\State EV $i$ computes the summation of the received messages $v_i^{\ell}$ defined in \eqref{12s} and broadcasts itno p to other agents. 

\State Each EV collects in total $n$ points, i.e., $(\alpha_l,v_l^{\ell}),\forall l=1,\ldots,n$, then reconstructs the constant term of the polynomial in 
\eqref{13s} based on the secret reconstruction method introduced in \eqref{recosntruction}.

\State  The $i$th EV updates the primal variable $\bm{x}_i^{\ell}$ by \eqref{11a} and the dual variable $\lambda_i^{\ell}$ by \eqref{11b}.

\State Each agent $i$ calculates the error $\epsilon_i^{\ell}$.

\State  $\ell=\ell+1$. 

\EndWhile
\end{algorithmic}
\label{protocol_1}
\end{protocol}

\noindent \textbf{Theorem 1:} (\textit{Correctness}) Within the precision range $\delta$, the results calculated by Protocol \ref{protocol_1} are exactly the same as those calculated without privacy preservation. \hfill $\blacksquare$

Theorem 1 states that the proposed privacy-preserving protocol does not jeopardise the accuracy of the PGA except for the rounding error introduced by real to integer number transformation. In fact, Protocol \ref{protocol_1} can achieve arbitrarily higher precision level by increasing the decimal digits $\delta$ at the cost of computating load.

\noindent \textbf{Theorem 2:} (\textit{Privacy-preservation against 
honest-but-curious agents and external eavesdroppers})
Privacy of all agents are guaranteed if the number of agents is larger than two and each agent carries out the proposed protocol accordingly. \hfill $\blacksquare$

Theorem 2 provides that the participants' privacy preservation can be guaranteed under Protocol \ref{protocol_1} against both  internal and external attacks. The internal attacks are launched by the \emph{honest-but-curious agent} who follows the protocol but may observe the intermediate data to infer the private information of other participants. The \emph{external eavesdropper} can launch external attacks by wiretapping and intercepting exchanged messages from the communication channels to invade the privacy of the agents. The detailed privacy analysis will be presented in Section \ref{Privacy_Analysis}. 

\noindent \textbf{Remark 2:} 
The polynomial in \eqref{13s} is of degree $k$, hence at least $k+1$ points are required to reconstruct the polynomial. According to Protocol \ref{protocol_1}, each agent can have access to $n$ points on the 2-D plane corresponding to the inputs and outputs of \eqref{13s}. To guarantee a successful secret reconstruction, the number of agents $n$ and the degree $k$ of the polynomials  need to satisfy $n \geq k+1$. Note that increasing the degree $k$ does not improve the security level of Protocol \ref{protocol_1} against honest-but-curious agents or external eavesdroppers. However, a larger $k$ could potentially prevent the information leakage caused by the collaboration between honest-but-curious agents. Please refer to Section \ref{collaboration} for the extended discussion on the collaboration between honest-but-curious agents.
\hfill $\square$

\noindent \textbf{Remark 3:} 
Protocol \ref{protocol_1} is not limited to the specific valley-filling optimization problem in \eqref{s1}. We use this EV charging control example to illustrate the grid-level applications. The proposed protocol can be applied to a variety of different areas, e.g., controlling heterogeneous GERs for grid-level services \cite{zhou2017incentive}, energy management optimization where the neighbors are required to exchange information \cite{rahbari2014incremental}, and the general cooperative optimization problem in \cite{huo_control_letter}.
\hfill $\square$

\subsection{Privacy Analysis}
\label{Privacy_Analysis}

\subsubsection{Security analysis against honest-but-curious agents}
An \emph{honest-but-curious agent} can use the private information belonging to itself, the common knowledge between all the agents, and the received messages from others to infer the privacy of certain agents. To analyze the attacks from honest-but-curious agents, we simulate its behaviours to prove that no private information can be obtained under the proposed protocol. Let EV $i$ be an honest-but-curious agent. Then during the $\ell$th iteration, EV $i$ can have access to
\begin{equation}
    \mathbb{I}_{hi}^{\ell} = \{\alpha_j,p_i^{\ell}(\alpha_j), v_j^{\ell}, \forall j=1,
\ldots,n,s_i^{\ell}, p_j^{\ell}(\alpha_i), \forall j \neq i \}, 
\end{equation}
where $\mathbb{I}_{hi}^{\ell}$ denotes the accessible information to EV $i$. Suppose EV $i$ is interested in finding the secret $s_m^{\ell}$ that belongs to EV $m$. The information EV $i$ can use includes $\alpha_m$, $p_i^{\ell}(\alpha_m)$, and the message $p_m^{\ell}(\alpha_i)$ received from EV $m$. To reconstruct the secret $s_m^{\ell}$ contained in $p_m^{\ell}(\cdot)$, EV $i$ needs to obtain at least 
$k$ points on the 2-D plane. However, EV $i$ only has access to a single point $(\alpha_i,p_m^{\ell}(\alpha_i))$. Therefore, it is impossible for EV $i$ to reconstruct the secret $s_m^{\ell}$. Another useful information for EV $i$ is the collected points $(\alpha_i,v_l^{\ell}),\forall l=1,\ldots,n$, from which it can deduce the summation of secrets $\sum_{l=1}^{n} s_l^{\ell}$ from all the agents. However, when there exist other EVs other than EV $i$ and EV $m$, having the summation  $\sum_{l=1}^{n} s_l^{\ell}$ is still insufficient for EV $i$ to infer any private information of EV $m$.

We next consider a special case where only two EVs exist, i.e., EV $i$ and EV $m$, and EV $i$ tries to infer the secret $s_m^{\ell}$ that belongs to EV $m$. Since EV $i$ knows its own secret $s_i^{\ell}$ and the summation $\sum_{l=1}^{n} s_l^{\ell}$, $s_m^{\ell}$ can be easily compromised. We argue that this rarely happens in real EV charging control cases as well as other large-scale networked optimization problems because of the large agent population size. To address this concern, Theorem 2 states the requirement that more than two agents should be involved during the execution of Protocol \ref{protocol_1}.

\subsubsection{Security analysis against external eavesdroppers}

Another principal adversary is the  \emph{external eavesdropper}. For general SS-based schemes (presented in Section \ref{chapter_SS}), the shares are distributed by the manager, and $k$ shares can expose the secret. In those cases, an external eavesdropper can easily wiretap and collect all the transmitted messages, thus revealing the secret. We next simulate the behaviours of the external eavesdroppers and prove the security of Protocol \ref{protocol_1}.

 Suppose an external eavesdropper wiretaps all communication channels in Protocol \ref{protocol_1}. Then the external eavesdropper can have access to the following transmitted messages
\begin{equation}
    \mathbb{I}_e^{\ell} = \{p_i^{\ell}(\alpha_j), \forall i\neq j, v_i^{\ell}, \forall i,j=1,
\ldots,n \},
\label{18}
\end{equation}
where $\mathbb{I}_e^{\ell}$
denotes the accessible information to the external eavesdropper. The message $p_i^{\ell}(\alpha_i), i=1,
\ldots,n$, is kept private to the $i$th EV, therefore this message cannot be directly overheard by the external eavesdroppers. However, the external eavesdropper can deduce the value of $p_i^{\ell}(\alpha_i)$ by wiretapping the value of $v_i^{\ell}$  due to the summation operation in \eqref{12s}. Therefore, the external eavesdropper can have access to $\{p_i^{\ell}(\alpha_j), \forall i,j=1,\ldots,n\}$. In summary, the external eavesdropper can have access to the outputs of the polynomials in \eqref{11s}
and the polynomial in \eqref{13s}. In order to derive the secrets of the agents, the external eavesdropper also need to know the inputs of those polynomials. Instead of distributing the shares in the form of both the inputs and outputs of a polynomial, the proposed protocol keeps the inputs $\alpha_i,\forall i=1,\ldots,n$, private to the agents. Hence, any external eavesdropper cannot reconstruct agents' secrets with only outputs of those polynomials. 

\subsubsection{Collusion between honest-but-curious agents}\label{collaboration} Honest-but-curious agents may collude to infer the privacy of other agents. However, few research has addressed the collusion between honest-but-curious agents. In this paper we extend the privacy analysis against this type of attacks to achieve enhanced security. The degree of the polynomial in \eqref{11s} is $k$, hence at least $k$ honest-but-curious agents need to collude to infer the privacy of one specific agent. To prevent from this collusion, it is possible to increase the degree $k$ within the limit of $k \leq n-1$. By increasing $k$,  more honest-but-curious agents are required to work together in order to infer the privacy, which is more demanding for the honest-but-curious agents. Consequently, a higher security level towards colluding honest-but-curious agents can be achieved. 

Note that when $k$ is at the maximum, i.e., $k=n-1$, the received shares from all $n$ agents are needed to reconstruct a secret. In this scenario, all agents have to be honest-but-curious so as to retrieve the secret. However, it is trivial to discuss the case where all agents are honest-but-curious and trying to collaborate with each other, because they share information with each other through collusion and secrets do not exist.

\section{Simulation Results}

In this section, we conduct valley-filling simulations with 20 EVs for an apartment complex in the residential area. Assume this complex is equipped with a 20 level-2 electric vehicle supply equipment (maximum 6.6 kW). All EVs need to be charged to the desired levels by the end of the valley-filling period. The net demand profile (system demand minus wind and solar) was taken from California Independent System Operator on 07/21/2021 and 07/22/2021 \cite{CISO}. We scale the demand profile by 200 to simulate the daily residential load profile of the complex. The valley-filling starts at 19:00 in the evening and ends at 7:00 in the morning next day. Herein, we have in total 48 time slots by sampling every 15 minutes within 12 hours. The preserved decimal fraction digits is set as $\delta=3$. The primal step sizes are uniformly chosen as $\gamma_l = 0.01, \forall l=1,\dots,n$, and the dual step size is $\beta = 1$. For the $i$th EV, the input $\alpha_i$ is chosen to be $i$. Initial values of $\bm{x}_i$ and $\lambda_i$ are set to be zeros. The maximum charging rates of all EVs are uniformly set as  $r_u = 6.6$ kW. The charging demands $d_i$'s of the EVs are randomly distributed in $[10,20]$ kWh. The degree of all polynomials are set as $k=3$ and the integer field is chosen as $\mathbb{E} = [0, 2^{31}-1)$. \begin{figure}[!htbp]
\centering
 \includegraphics[width=0.5\textwidth,trim = 0mm 0mm 0mm 10mm, clip]{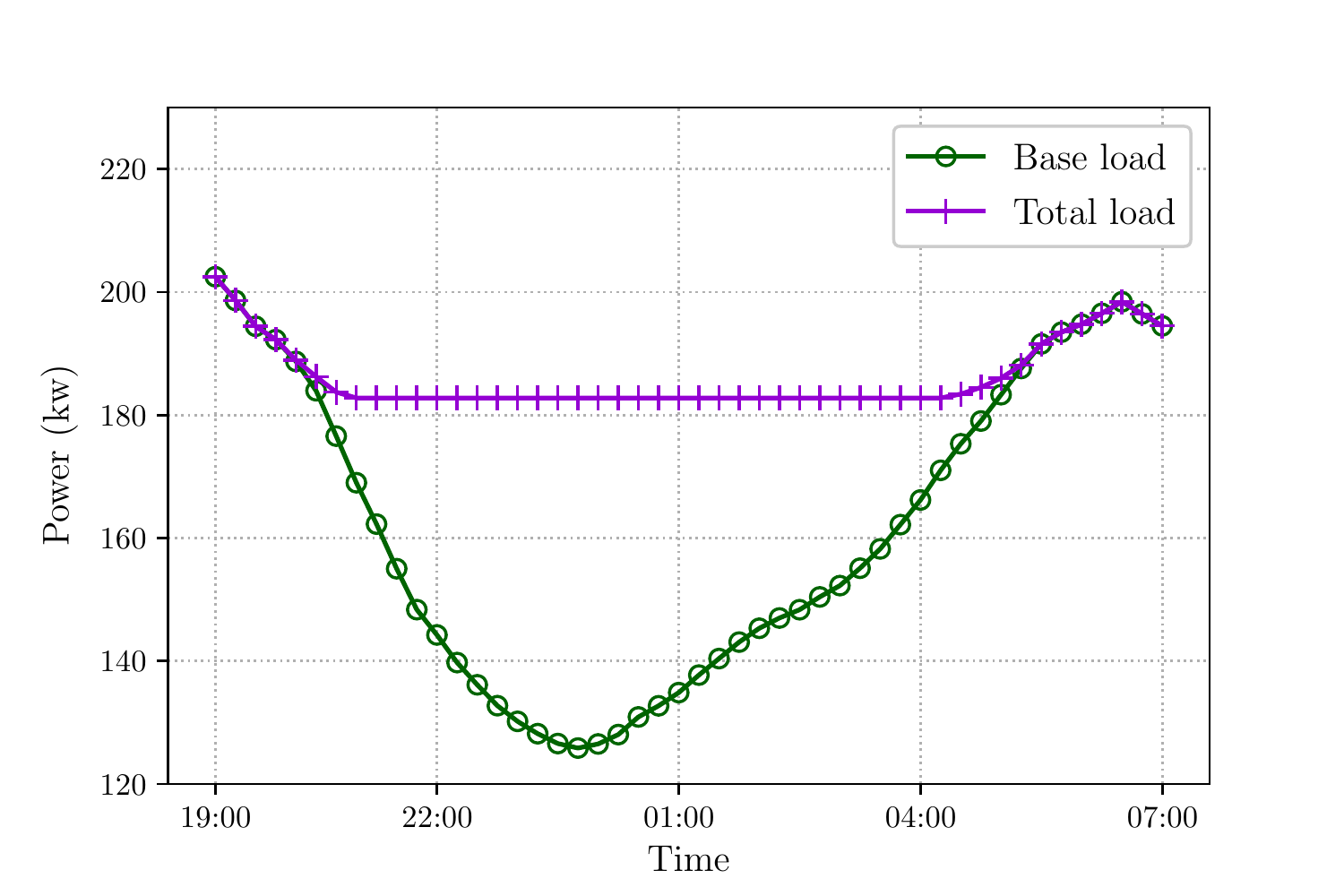}
    \caption{The baseline load profile and the total load profile under the proposed  privacy-preserving EV charging control protocol.}
    \label{valley_filling}
\end{figure}

\begin{figure}[!htbp]
\vspace*{-2mm}
    \centering
 \includegraphics[width=0.5\textwidth,trim = 0mm 0mm 0mm 10mm, clip]{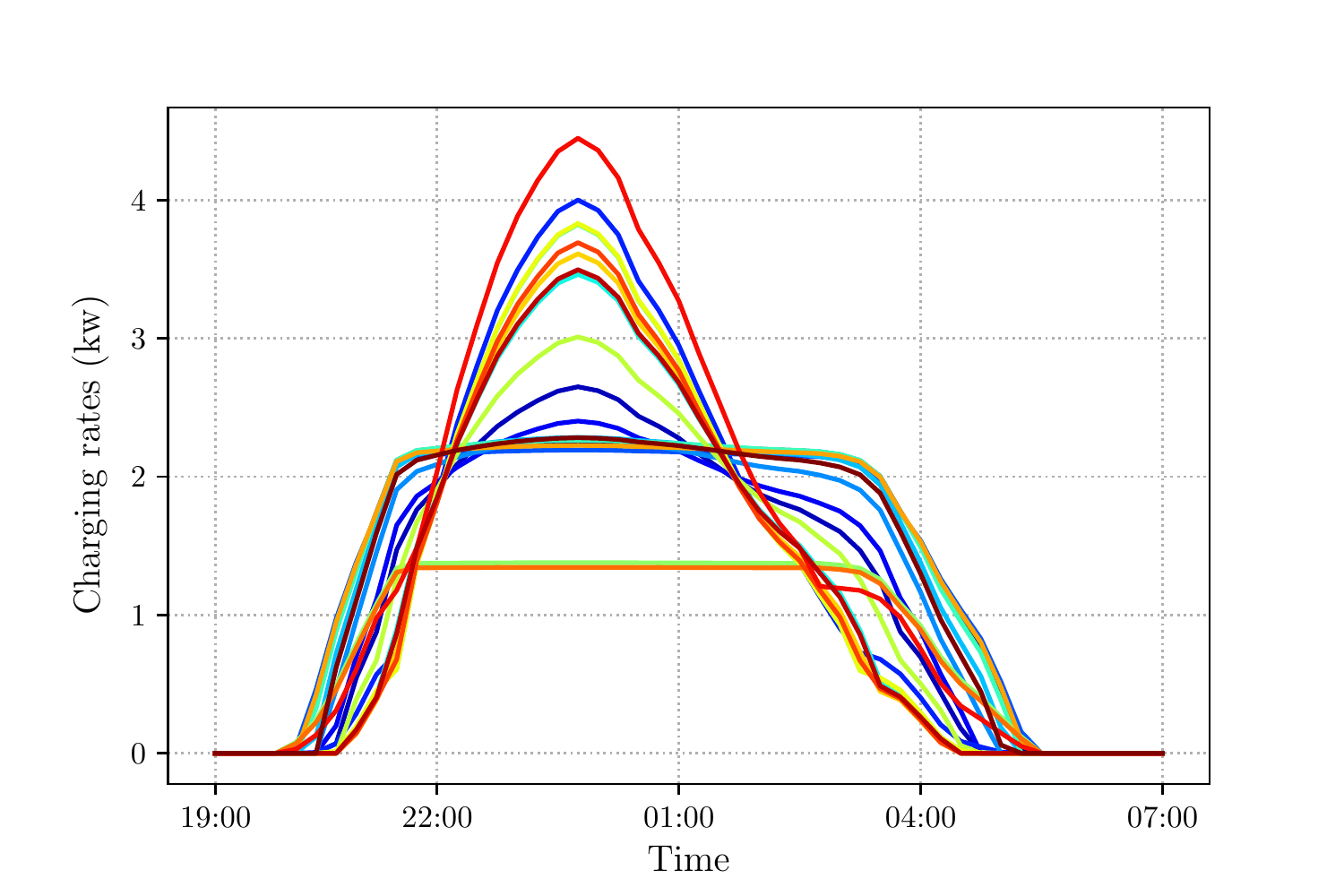}
    \caption{Charging profiles of all EVs after 300 iterations during the valley-filling period.}
    \label{charging_profiles}
\end{figure}

\begin{figure}[!htbp]
\vspace*{-2mm}
    \centering
 \includegraphics[width=0.5\textwidth,trim = 0mm 0mm 0mm 7mm, clip]{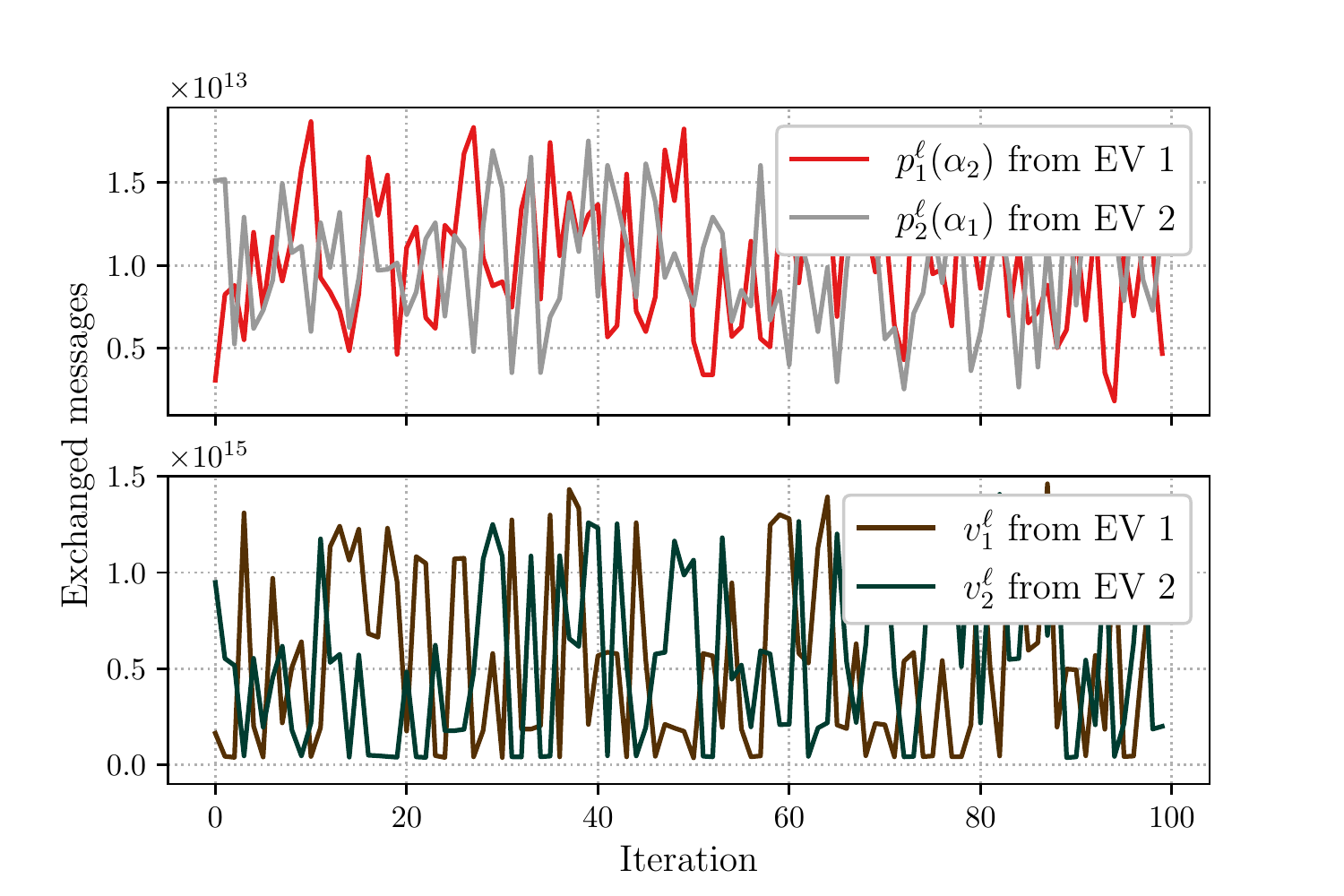}
    \caption{Messages sent from  EV 1 and EV 2.}
    \label{exchanged_messages}
\end{figure}

\begin{figure}[!htbp]
    \centering
 \includegraphics[width=0.5\textwidth,trim = 0mm 0mm 0mm 7mm, clip]{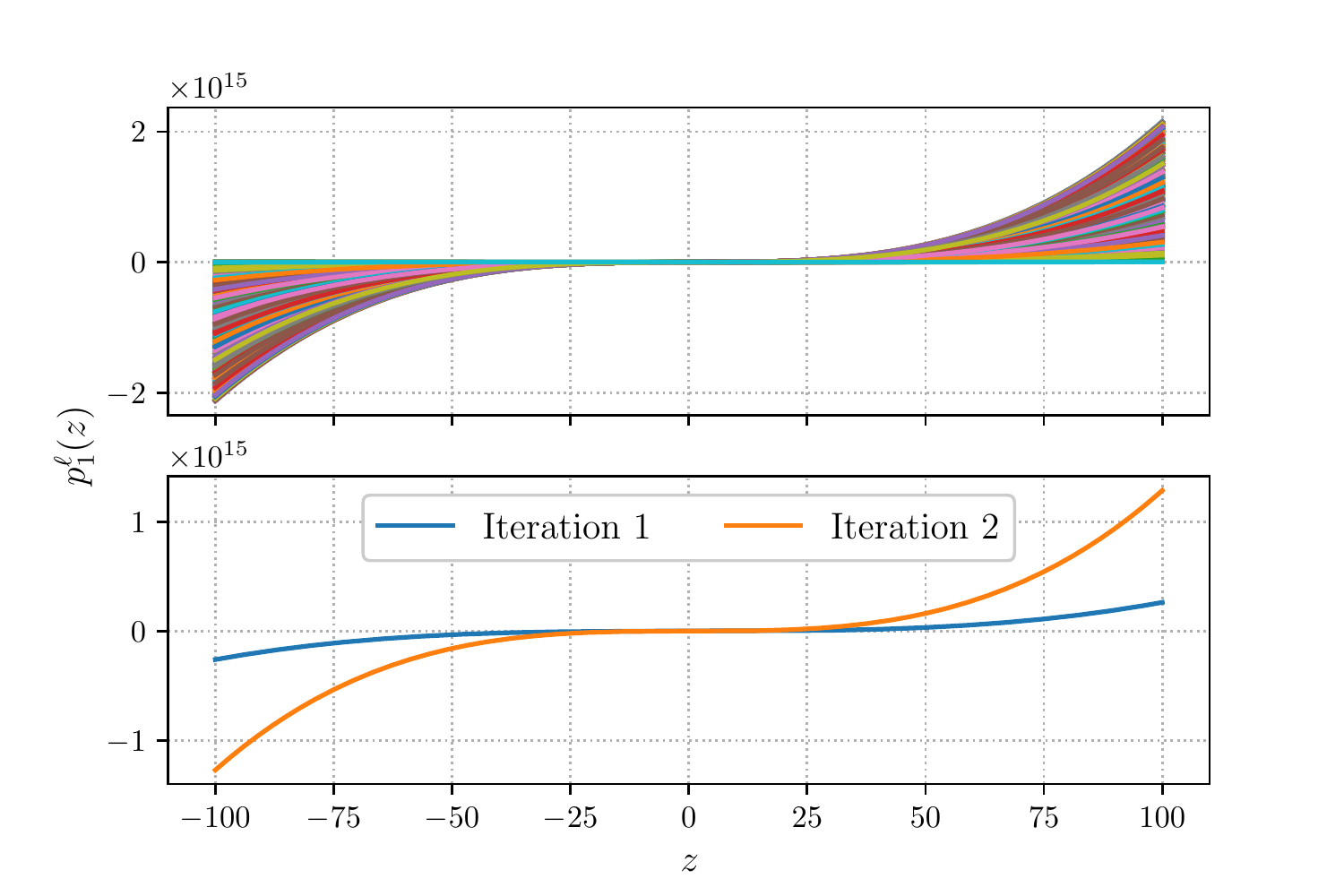}
    \caption{Polynomials generated by EV 1 \emph{w.r.t} the secret $\bm{x}_1(1)$.}
    \label{polynomials}
\end{figure}

Fig. \ref{valley_filling} presents the valley-filling results where the baseline load is flattened into the total load by controlling the EVs' charging loads. The final charging profiles of the 20 EVs are shown in 
Fig. \ref{charging_profiles}. All the EVs charge at the fastest rates ranging from 1 kW to 4.5 kW at  around $00:00$ to fill the midnight deep valley. Without the loss of generality, we  present the messages sent by two EVs in Fig. \ref{exchanged_messages}, i.e., $p_1^{\ell}(\alpha_2)$ and $v_1^{\ell}$ sent by EV 1, and $p_2^{\ell}(\alpha_1)$ and $v_2^{\ell}$ sent by EV 2 within 100 iterations. As can be readily observed, all the transmitted messages are randomized to achieve privacy preservation. Fig. \ref{polynomials} gives the polynomials generated by EV 1 that are of degree 3 and aim at randomizing and protecting the first element of the decision variable $\bm{x}_i$. In total 300 polynomials $p_1^{\ell}(z),\ell=1,\ldots,300$ are presented in the upper plot of Fig. \ref{polynomials}, and the polynomials from the first two iterations are shown in the lower plot for clear presentation.  All other coefficients, i.e., $c_{11}$, $c_{12}$, and $c_{13}$, are randomized and vary in each iteration as well.

\section{Conclusion and Future Research}

In this paper, we proposed a novel distributed  privacy-preserving EV charging control protocol. We integrated SS into PGA to secure the privacy of the EVs against individual and colluding honest-bust-curious agents as well as external eavesdroppers. Through theoretical security analyses, privacy of all EVs are guaranteed against those adversaries. The proposed protocol enjoys lower computing cost compared to HE-based methods and achieves scalability through its distributed setting. Simulation results of a valley-filling problem show the efficiency and efficacy of the proposed privacy-preserving protocol. Some limitations of the proposed method were identified and those limitations motivate future research directions including decreasing the communication loads between the participants and expanding the proposed protocol to generic optimization problems in other research fields.

\section*{Acknowledgement}
This work was supported in partial by US Department of Energy DE-EE0009658. 



\bibliographystyle{IEEEtran}

\bibliography{bibliography}


\end{document}